\documentclass[12pt]{article}
\usepackage{amsmath}
\usepackage{amssymb}
\usepackage{amsfonts}
\usepackage{latexsym}
\usepackage{color}
\usepackage{float}
\usepackage{tikz}
\usepackage{graphicx,graphics,epstopdf}
\usepackage{amscd}
\usepackage{subfigure}
 \newtheorem{thm}{Theorem}[section]

\newcommand{\mc}{\mathcal}
\newcommand{\Hil}{\mc H}
\newcommand{\D}{{\mc D}}

\newcommand{\1}{1 \!\! 1}
\newcommand{\be}{\begin{equation}}
\newcommand{\en}{\end{equation}}
\newcommand{\bea}{\begin{eqnarray}}
\newcommand{\ena}{\end{eqnarray}}
\newcommand{\beano}{\begin{eqnarray*}}
\newcommand{\enano}{\end{eqnarray*}}
\newcommand{\bee}{\begin{enumerate}}
\newcommand{\ene}{\end{enumerate}}

\newcommand{\Sc}{{\mathcal S}}

\newcommand{\F}{{\mathcal F}}
\newcommand{\G}{{\mathcal G}}

\newcommand{\Lc}{{\mathcal L}}

\newcommand{\U}{{\mathcal U}}
\newcommand{\V}{{\mathcal V}}

\newcommand\blfootnote[1]{%
  \begingroup
  \renewcommand\thefootnote{}\footnote{#1}%
  \addtocounter{footnote}{-1}%
  \endgroup
}

\begin{document}

\thispagestyle{empty}

\vspace*{2cm}

\begin{center}
{\Large \bf {Two-dimensional non commutative  Swanson model and its bicoherent states}}   \vspace{2cm}\\

{\large F. Bagarello$^{1,2,3}$,F.Gargano$^{1}$, S. Spagnolo$^{1}$}\blfootnote{\hspace*{-0.75cm} 1.  DEIM -Dipartimento di Energia, ingegneria dell' Informazione e modelli Matematici, Scuola Politecnica, Universit\`a di Palermo, I-90128  Palermo, Italy\\
2. INFN, Sezione di Napoli\\
3. Department of Mathematics and Applied Mathematics, University of Cape Town, South Africa\\

E-mail: fabio.bagarello@unipa.it (FB), francesco.gargano@unipa.it (FG), salvatore.spagnolo@unipa.it (SS)}
\end{center}

\begin{abstract}
We introduce an extended version of the Swanson model, defined on a two-dimensional non commutative space, which can be
diagonalized exactly by making use of pseudo-bosonic operators. Its eigenvalues are explicitly computed  and the biorthogonal sets of eigenstates of the Hamiltonian and of its adjoint are explicitly constructed. We also show that it is possible to construct two displacement-like operators from which a family of bi-coherent states can be obtained. These states are shown to be eigenstates of the deformed lowering operators, and their  projector allows to produce a suitable resolution of the identity in a dense subspace of $\Lc^2(\Bbb R^2)$.

% We show that despite the fact that these sets are complete and biorthogonal, they involved an unbounded metric operator and therefore do not constitute (Riesz) bases for the Hilbert space $\Lc^2(\Bbb R^2)$, but instead only D-quasi bases.
\end{abstract}

\newpage

\section{Introduction}

In the past twenty years or so a lot of interest arose on the so-called $PT$-quantum mechanics. This was mainly due to the paper in \cite{ben} where the authors introduced a manifestly non self-adjoint, but $PT$-symmetric, Hamiltonian with purely real (and discrete) eigenvalues. Here $P$ and $T$ are the parity and the time-reversal operators. The main point was that having a physical, rather than a mathematical, condition which guarantees the reality of the spectrum would be quite interesting and more natural for the physicists community.
One of the very famous examples of this situation was later introduced in \cite{swan}, with the Hamiltonian
$$
H_\nu=\frac{1}{2}\left(p^2+x^2\right)-\frac{i}{2}\,\tan(2\nu)\left(p^2-x^2\right).
$$
{Here $\nu$ is a real parameter taking value in $\left(-\frac{\pi}{4},\frac{\pi}{4}\right)\setminus\{0\}$. This model has relevant mathematical and physical implications, and was discussed in terms of the so-called $\D$-pseudo bosons, \cite{bag1,baginbagbook}}. Here we consider a two dimensional version of this model living in a non-commutative plane, and we show how this model can again be understood in terms of pseudo-bosons. Also, we briefly discuss what changes if we do not assume $\nu$ to be strictly real. Finally, we construct bicoherent states associated to the model and we check some of their properties.

\section{Non-commutative two dimensional harmonic oscillator with linear
terms}

The Hamiltonian we want to consider here, depending on two parameters $\nu$ and $\theta$, is the following
$$
H_{\nu,\theta}=\frac{1}{2\cos(2\nu)}\,\left\{\hat p_1^2\left(e^{-2i\nu}+\frac{\theta^2}{4}\,e^{2i\nu}\right)+\hat x_1^2e^{2i\nu}+\hat p_2^2\left(e^{-2i\nu}+\frac{\theta^2}{4}\,e^{2i\nu}\right)+\right.$$
\be
\left. +\hat x_2^2e^{2i\nu}+2\theta\left(\hat x_1\hat p_2-\hat x_2\hat p_1\right)\right\},
\label{21}\en
where the operators $\hat x_j$ and $\hat p_j$ satisfy the following commutation rules:
\be
[\hat x_j,\hat p_k]=i\delta_{j,k}\1,\qquad [\hat x_1,\hat x_2]=i\theta\1,\qquad [\hat p_j,\hat p_k]=0.
\label{22}\en
{The Hamiltonian $H_{\nu,\theta}$ can be seen as a reasonable two dimensional version of the one dimensional Swanson model discussed in \cite{swan,dapro,bag1}, defined in a non commutative two-dimensional plane. Here $\theta$ is the {\em non-commutativity parameter}, while $\nu$ is a real\footnote{In some part of the paper we will remove the assumption of $\nu$ being real, and see what happens when we put an imaginary part in it.} {\em non self-adjointness parameter}, taking values in $I:=\left(-\frac{\pi}{4},\frac{\pi}{4}\right)$.}  Whenever $\nu\in I$ is not zero, $H_{\nu,\theta}\neq H_{\nu,\theta}^\dagger$. On the other hand, $H_{\nu=0,\theta}=H^\dagger_{\nu=0,\theta}$. Moreover, if we take $\theta=0$, i.e. if we go back to a commuting plane, we see that
$$
H_{\nu,\theta}=\frac{1}{2\cos(2\nu)}\,\left\{\hat p_1^2e^{-2i\nu}+\hat x_1^2e^{2i\nu}+\hat p_2^2e^{-2i\nu}+\hat x_2^2e^{2i\nu}\right\},
$$
which is exactly the two dimensional version of the Hamiltonian considered in \cite{dapro,bag1}: removing the non-commutativity (by sending $\theta$ to zero) returns the standard Swanson model, in two dimensions and without interactions. Finally, if we take $\nu=\theta=0$, $H_{\nu,\theta}$ is nothing but the Hamiltonian of a two-dimensional harmonic oscillator.

\vspace{2mm}

Despite of its apparently complicated expression, the operator $H_{\nu,\theta}$ can be diagonalized in a rather simple way, by making use of the $\D$-pseudo bosons introduced by one of us (F.B.), and widely analyzed in \cite{baginbagbook}. In fact, let:
\be\left\{
\begin{array}{ll}
A_1=\frac{1}{\sqrt{2}}\left(\hat x_1 e^{i\nu}+\frac{\theta}{2}\,\hat p_2e^{i\nu}+i\hat p_1e^{-i\nu}\right),\quad A_2=\frac{1}{\sqrt{2}}\left(\hat x_2 e^{i\nu}-\frac{\theta}{2}\,\hat p_1e^{i\nu}+i\hat p_2e^{-i\nu}\right)\\
B_1=\frac{1}{\sqrt{2}}\left(\hat x_1 e^{i\nu}+\frac{\theta}{2}\,\hat p_2e^{i\nu}-i\hat p_1e^{-i\nu}\right),\quad B_2=\frac{1}{\sqrt{2}}\left(\hat x_2 e^{i\nu}-\frac{\theta}{2}\,\hat p_1e^{i\nu}-i\hat p_2e^{-i\nu}\right).
\end{array}
\right.
\label{23}\en
First of all, it is clear that, for $\nu\neq0$, $B_j\neq A_j^\dagger$, $j=1,2$. Moreover, it is easy to check using (\ref{22}) that \be
[A_j,B_k]=\delta_{j,k}\1,\qquad [A_j,A_k]=[B_j,B_k]=0.
\label{24}\en
Then these operators satisfy the two dimensional pseudo-bosonic rules, \cite{baginbagbook}. More important, in terms of them our Hamiltonian $H_{\nu,\theta}$ in (\ref{21}) acquires a much simpler form:
\be
H_{\nu,\theta}=\frac{1}{\cos(2\nu)}\,\left(B_1A_1+B_2A_2+\1\right),
\label{25}\en
which is manifestly non self-adjoint for $\nu\neq0$. Indeed we have \be H_{\nu,\theta}^\dagger=\frac{1}{\cos(2\nu)}\,\left(A_1^\dagger B_1^\dagger +A_2^\dagger B_2^\dagger+\1\right),\label{25bis}\en
which is different from $H_{\nu,\theta}$  when $\nu\neq0$.

Once the Hamiltonian has been written as in (\ref{25}), we can use the general settings described in details in \cite{baginbagbook}: we have to look first for the vacua $\varphi_{0,0}$ and $\Psi_{0,0}$ of $A_j$ and $B_j^\dagger$, $j=1,2$, and identify a set $\D$, dense in the Hilbert space, such that $\varphi_{0,0},\Psi_{0,0}\in\D$ and $\D$ is left stable under the action of $A_j$,
 $B_j$ and their adjoints. Then, we act on $\varphi_{0,0}$ and $\Psi_{0,0}$  with $B_j$ and $A_j^\dagger$, respectively, producing two biorthogonal sets of eigenstates of $H_{\nu,\theta}$ and $H_{\nu,\theta}^\dagger$. The procedure here is particularly simple if we adopt the  so-called Bopp shift to represent the commutation rules in (\ref{22}). In fact, let us introduce two pairs of self-adjoint operators $(x_j,p_j)$, $j=1,2$, satisfying $[x_j,p_k]=i\delta_{j,k}\1$, $[x_j,x_k]=[p_j,p_k]=0$. Then (\ref{22}) are recovered if we assume that
\be \hat x_1= x_1-\frac{\theta}{2}\,
p_2,\quad\hat x_2= x_2+\frac{\theta}{2}\,  p_1,\quad \hat p_1= p_1,\quad \hat p_2= p_2. \label{26}\en
In terms of these operators $A_j$ and $B_j$ can be rewritten as
\be\left\{
\begin{array}{ll}
A_1=\frac{1}{\sqrt{2}}\left(x_1 e^{i\nu}+e^{-i\nu}\frac{d}{dx_1}\right),\quad
A_2=\frac{1}{\sqrt{2}}\left(x_2 e^{i\nu}+e^{-i\nu}\frac{d}{dx_2}\right)\\
B_1=\frac{1}{\sqrt{2}}\left(x_1 e^{i\nu}-e^{-i\nu}\frac{d}{dx_1}\right),\quad
B_2=\frac{1}{\sqrt{2}}\left(x_2 e^{i\nu}-e^{-i\nu}\frac{d}{dx_2}\right),
\end{array}
\right.
\label{27}\en
which shows that, in terms of $(x_j,p_j)$, the two pairs $(A_1,B_1)$ and $(A_2,B_2)$ are completely independent. Hence, the construction of the set of eigenvectors of $H_{\nu,\theta}$, $\F_\varphi=\{\varphi_{n_1,n_2}(x_1,x_2)\}$, and the set of eigenvectors of $H_{\nu,\theta}^\dagger$, $\F_\Psi=\{\Psi_{n_1,n_2}(x_1,x_2)\}$, can be carried out simply considering tensor products of the one dimensional construction already considered in for instance in \cite{bag1,BGV15}.
In particular, the two vacua of $A_j$ and $B_j^\dagger$ are easily found:
\beano
\varphi_{0,0}(x_1,x_2)=\varphi_{0}(x_1)\varphi_{0}(x_2)&=&N_1
\exp\left\{-\frac{1}{2}\,e^{2i\nu}\,(x_1^2+x_2^2)\right\},
\\
\Psi_{0,0}(x_1,x_2)=\Psi_{0}(x_1)\Psi_{0}(x_2)&=&N_2 \exp\left\{-\frac{1}{2}\,e^{-2i\nu}\,(x_1^2+x_2^2)\right\},
\enano
where $N_1$ and $N_2$ are  normalization constants satisfying $\overline{N_1}N_2=\frac{e^{2i\nu}}{\pi}$, to ensure that $\left<\varphi_{0,0},\Psi_{0,0}\right>=1$.

Notice that, since $\Re(e^{\pm 2i\nu})=\cos(2\nu)>0$ for all $\nu\in I$, both $\varphi_{0,0}(x_1,x_2)$ and $\Psi_{0,0}(x_1,x_2)$
belong to $\Sc({\Bbb R}^2)$, and therefore to $\Lc^2({\Bbb R}^2)$. Now, if we define
\beano
\varphi_{n_1,n_2}&=&\frac{1}{\sqrt{n_1!n_2!}}B_1^{n_1}B_2^{n_2}\varphi_{0,0},\\
\Psi_{n_1,n_2}&=&\frac{1}{\sqrt{n_1!n_2!}}(A_1^\dag)^{n_1}(A_2^\dag)^{n_2}\Psi_{0,0},
\enano
we get, see \cite{bag1},
\bea
\left\{
\begin{array}{ll}
\varphi_{n_1,n_2}(x_1,x_2)=\frac{N_1}{\sqrt{2^{n_1+n_2}\,n_1!\,n_2!}}
\,H_{n_1}\left(e^{i\nu}x_1\right)\,H_{n_2}\left(e^{i\nu}x_2\right)\times\\
\qquad\qquad\times\exp\left\{-\frac{1}{2}\,e^{2i\nu}\,(x_1^2+x_2^2)\right\}, \label{2a} \\
\Psi_{n_1,n_2}(x_1,x_2)=\frac{N_2}{\sqrt{2^{n_1+n_2}\,n_1!\,n_2!}}
\,H_{n_1}\left(e^{-i\nu}x_1\right)\,H_{n_2}\left(e^{ -i\nu}x_2\right)\times\\
\qquad\qquad\times\exp\left\{-\frac{1}{2}\,e^{-2i\nu}\,(x_1^2+x_2^2)\right\},
\end{array}
\right.
\ena
where $H_n(x)$ is the n-th Hermite polynomial. We see from these formulas that, for all $n_j\geq0$,
$\frac{1}{N_1}\,\varphi_{n_1,n_2}(x_1,x_2)$ coincides with $\frac{1}{N_2}\,\Psi_{n_1,n_2}(x_1,x_2)$, with $\nu$ replaced by $-\nu$. Moreover, they all belong to $\Sc({\Bbb R}^2)$, and therefore to $\Lc^2({\Bbb R}^2)$, which is a clear
indication that $\varphi_{0,0}(x_1,x_2)\in D^\infty(B_j)$ and $\Psi_{0,0}(x_1,x_2)\in D^\infty(A_j^\dagger)$, $j=1,2$. Also, they are biorthogonal $\left<\varphi_{n_1,n_2},\Psi_{m_1,m_2}\right>=\delta_{n_1,m_1}\delta_{n_2,m_2}$, for all $n_j, m_j\geq0$, and the following equations are satisfied:
\bea\label{pbn}
\left\{
\begin{array}{ll}
A_1\varphi_{n_1,n_2}=\sqrt{n_1}\varphi_{n_1-1,n_2}, \quad A_2\varphi_{n_1,n_2-1}=\sqrt{n_2}\varphi_{n_1,n_2-1},\\
B_1^\dag\Psi_{n_1,n_2}=\sqrt{n_1}\Psi_{n_1-1,n_2}, \quad B_2^\dag\Psi_{n_1,n_2}=\sqrt{n_2}\Psi_{n_1,n_2-1},\\
B_1A_1\varphi_{n_1,n_2}=n_1\varphi_{n_1,n_2},\quad B_2A_2\varphi_{n_1,n_2}=n_2\varphi_{n_1,n_2},\\
(B_1A_1)^\dag\Psi_{n_1,n_2}=n_1\Psi_{n_1,n_2},\quad (B_2A_2)^\dag\Psi_{n_1,n_2}=n_2\Psi_{n_1,n_2}.\\
\end{array}
\right.
\ena

Following the same arguments as in \cite{baginbagbook}, it is possible to check that the norm of these vectors,  $\left\|\varphi_{n_1,n_2}\right\|$ and $\left\|\Psi_{n_1,n_2}\right\|$, diverge with $n_j$. Then,  $\F_\varphi$ and
$\F_\Psi$ are not Riesz bases, and not even bases. We are still left with the possibility that they are $\G$-quasi bases, for a
suitable set $\G$ dense in $\Lc^2({\Bbb R}^2)$, see below. Indeed, this is the case, as we can check extending, once again, what was done in \cite{baginbagbook} in the one-dimensional case. We don't give the details here, since they do not differ significantly from what is done in \cite{bag1,baginbagbook}. We only  stress that the crucial ingredient is provided by the  operator
$$
T_{\nu}=e^{i\,\frac{\nu}{2}\,\left(x_1\,\frac{d}{dx_1}+\frac{d}{dx_1}\,x_1\right)}e^{i\,\frac{\nu}{2}\,\left(x_2\,\frac{d}{dx_2}+\frac{d}{dx_2}\,x_2\right)},
$$
which maps (except for a normalization constant) the orthonormal basis of a two-dimensional harmonic oscillator, $\F_e$, into $\F_\varphi$. In the same way $(T^{-1})^\dagger$ maps (again, except for a normalization constant) the same basis into $\F_\Psi$. Then, calling $\D_e$ the linear span of $\F_e$ , which is obviously dense in $\Lc^2(\Bbb R^2)$, it turns out that $\F_\varphi$ and  $\F_\Psi$ are $\D_e$-quasi bases. This means that, for all $f,g\in\D_e$, the following resolution of the identity holds true:
\be
\left<f,g\right>=\sum_{n_1,n_2}\left<f,\varphi_{n_1,n_2}\right>\left<\Psi_{n_1,n_2},g\right>=
\sum_{n_1,n_2}\left<f,\Psi_{n_1,n_2}\right>\left<\varphi_{n_1,n_2},g\right>
\label{2add1}\en
Notice that both $T_{\nu}$ and $T_{\nu}^{-1}$ are unbounded. This can be easily understood easily, since both these operators are not everywhere defined on $\Lc^2({\Bbb R}^2)$.

Finally, the metric operator can now be explicitly deduced:  $\Theta:=\frac{1}{\pi\,|N_1|^2}\,T_{\nu}^{-2}$, which is unbounded, with unbounded inverse. Moreover, $(A_j, B^\dagger_j)$ are $\Theta$-conjugate in the sense of \cite{bag2}, and $H_{\nu,\theta}$ is similar to a self-adjoint Hamiltonian: $h_{\nu,\theta} f=T_{\nu}\, H_{\nu,\theta}\, T_{\nu}^{-1} f$, for all $f$ in a suitable dense domain of $\Lc^2({\Bbb R}^2)$,
where $h_{\nu,\theta}=\frac{1}{\cos{2\nu}}\left(a_1^\dagger a_1+a_2^\dagger a_2+\1\right)$, $a_j=\frac{1}{\sqrt{2}}(x_j+ip_j)$.

\section{Bi-coherent states}

We now consider the two pairs  of pseudo-bosonic operators $(A_j,B_j),\quad j=1,2$, behaving as in the  previous Section,
in order to construct a generalized version of the canonical coherent states.
First of all we introduce $\forall z,w\in\mathbb C$ the two displacement-like operators
\bea
\U(z,w)=e^{zB_1-\bar{z}A_1}e^{wB_2-\bar{w}A_2},\quad \V(z,w)=e^{zA_1^\dag-\bar{z}B_1^\dag}e^{wA_2^\dag-\bar{w}B_2^\dag}.
\ena
Of course these operators are not unitary and they are possibly not even bounded. Hence, at the moment, they should be understood as formal objects.

If we assume that the Baker-Campbell-Hausdorff relation can be applied to $\U(z,w)$ and $\V(z,w)$, due to the commutation relations $[A_j,[A_j,B_j]]=[B_j,[A_j,B_j]]=0$, $j=1,2$, we obtain the following alternative representations:
\bea
\nonumber\U(z,w)&=&e^{-\frac{|z|^2+|w|^2}{2}}e^{zB_1}e^{-\bar{z}A_1}e^{wB_2}e^{-\bar{w}A_2}=e^{\frac{|z|^2+|w|^2}{2}}e^{-\bar{z}A_1}e^{zB_1}e^{-\bar{w}A_2}e^{wB_2}\\ \nonumber\V(z,w)&=&e^{-\frac{|z|^2+|w|^2}{2}}e^{zA_1^\dag}e^{-\bar{z}B_1^\dag}e^{wA_2^\dag}e^{-\bar{w}B_2^\dag}=e^{\frac{|z|^2+|w|^2}{2}}e^{-\bar{z}B_1^\dag}e^{zA_1^\dag}e^{-\bar{w}B_2^\dag}e^{wA_2^\dag},\\
\label{factV}
\ena
so that
\bea
\nonumber\U(z,w)^{-1}=\U(-z,-w)=\V(z,w)^\dag,\quad \V(z,w)^{-1}=\V(-z,-w)=\U(z,w)^\dag,
\ena
Now, bi-coherent states could be constructed in the following way:
\bea
\varphi(z,w)=\U(z,w)\varphi_{0,0},\quad \Psi(z,w)=\V(z,w)\Psi_{0,0},\label{BCS}
\ena
where $\varphi_{0,0},\Psi_{0,0}$ are the two vacua introduced in the previous section.
However, it is more convenient to define $\varphi(z,w)$ and $\Psi(z,w)$ via the following series representations:
 \bea
 \varphi(z,w)&=&e^{-\frac{|z|^2+|w|^2}{2}}\sum_{n_1,n_2\geq0}\frac{z^m w^n}{\sqrt{n_1!n_2!}}\varphi_{n_1,n_2},\label{vphiBCS}\\ \Psi(z,w)&=&e^{-\frac{|z|^2+|w|^2}{2}}\sum_{n_1,n_2\geq0}\frac{z^m w^n}{\sqrt{n_1!n_2!}}\Psi_{n_1,n_2}\label{vpsiBCS}.
 \ena
This is because, if we are able to prove that the series converge, then we don't need to take care of all the mathematical subtleties appearing if $\U(z,w)$ and $\V(z,w)$ are unbounded. On the other hand, it is not hard to prove that the above series converge $\forall z,w\in\mathbb{C}$, and that the states they define have interesting properties. For that, it is convenient to prove first a rather general result on bi-coherent states, which in a sense unifies and extend the results described in many papers recently, \cite{fb1}-\cite{fb5}.

\subsection{A general theorem}

Here we work with two biorthogonal families of vectors, $\F_\varphi=\{\varphi_n, \, n\geq0\}$ and $\F_\Psi=\{\Psi_n, \, n\geq0\}$ which are $\D$
-quasi bases for some dense subset of $\Hil$, see (\ref{2add1}). Consider an increasing sequence of real numbers $\alpha_n$ satisfying the  inequalities $0=\alpha_0<\alpha_1<\alpha_2<\ldots$. We call $\overline\alpha$ the limit of $\alpha_n$ for $n$ diverging, which coincides with $\sup_n\alpha_n$. We further consider two operators, $a$ and $b^\dagger$, which act as lowering operators respectively on $\F_\varphi$ and $\F_\Psi$ in the following way:
\be
a\,\varphi_n=\alpha_n\varphi_{n-1}, \qquad b^\dagger\,\Psi_n=\alpha_n\Psi_{n-1},
\label{31}\en
for all $n\geq1$, with $a\,\varphi_0=b^\dagger\,\Psi_0=0$.

\begin{thm}\label{theo1}
Assume that four strictly positive constants $A_\varphi$, $A_\Psi$, $r_\varphi$ and $r_\Psi$ exist, together with two strictly positive sequences $M_n(\varphi)$ and $M_n(\Psi)$ for which
\be
\lim_{n\rightarrow\infty}\frac{M_n(\varphi)}{M_{n+1}(\varphi)}=M(\varphi), \qquad \lim_{n\rightarrow\infty}\frac{M_n(\Psi)}{M_{n+1}(\Psi)}=M(\Psi),
\label{30}\en
where $M(\varphi)$ and $M(\Psi)$ could be infinity, such that, for all $n\geq0$,
\be
\|\varphi_n\|\leq A_\varphi\,r_\varphi^n M_n(\varphi), \qquad \|\Psi_n\|\leq A_\Psi\,r_\Psi^n M_n(\Psi).
\label{31b}\en
Then the following series:
\be
N(|z|)=\left(\sum_{k=0}^\infty\frac{|z|^{2k}}{(\alpha_k!)^2}\right)^{-1/2},
\label{32}\en
\be
\varphi(z)=N(|z|)\sum_{k=0}^\infty\frac{z^k}{\alpha_k!}\varphi_k,\qquad \Psi(z)=N(|z|)\sum_{k=0}^\infty\frac{z^k}{\alpha_k!}\Psi_k,
\label{33}\en
are all convergent inside the circle $C_\rho(0)$ centered in the origin of the complex plane and of radius $\rho=\overline\alpha\min\left(1,\frac{M(\varphi)}{r_\varphi},\frac{M(\Psi)}{r_\Psi}\right)$ . Moreover, for all $z\in C_\rho(0)$,
\be
a\varphi(z)=z\varphi(z), \qquad b^\dagger \Psi(z)=z\Psi(z).
\label{34}\en
Suppose further that a measure $d\lambda(r)$ does exist such that
\be
\int_0^\rho d\lambda(r) r^{2k}=\frac{(\alpha_k!)^2}{2\pi},
\label{35}\en
for all $k\geq0$. Then, for all $f,g\in\D$, calling $d\nu(z,\overline z)=d\lambda(r)d\theta$, we have
\be
\int_{C_\rho(0)}N(|z|)^{-2}\left<f,\Psi(z)\right>\left<\varphi(z),g\right>d\nu(z,\overline z)=$$
$$=\int_{C_\rho(0)}N(|z|)^{-2}\left<f,\varphi(z)\right>\left<\Psi(z),g\right>d\nu(z,\overline z)=
\left<f,g\right>
\label{36}\en

\end{thm}

The proof of the theorem is simple and will not be given here. Rather than this, there are few comments which are in order: first of all, we see from (\ref{31b}) that the norms of the vectors $\varphi_n$ and $\Psi_n$ need not being uniformly bounded, as it happened to be in \cite{fbbial}. On the contrary, they can diverge rather fastly with $n$. To see this, we juct consider $r_\varphi, r_\Psi>1$ and $M_n(\varphi)$ and $M_n(\Psi)$ constant sequences.

To apply the above theorem to the Swanson model we need to construct a two-dimensional version of it. This can be done in a natural way: suppose again we have two biorthogonal families of vectors, $\F_\varphi=\{\varphi_{n_1,n_2}, \, n_j\geq0\}$ and $\F_\Psi=\{\Psi_{n_1,n_2}, \, n_j\geq0\}$ which are $\D$
-quasi bases for some dense subset of $\Hil$. As we can see, these vectors depend on two sequences of natural numbers. Let now $\{\alpha_n\}$ and $\{\beta_n\}$ be two sequences of real numbers such that $0=\alpha_0<\alpha_1<\alpha_2<\ldots$ and $0=\beta_0<\beta_1<\beta_2<\ldots$. We call $\overline\alpha$ and $\overline\beta$ their limits. We further consider four operators, $a_j$ and $b_j^\dagger$, $j=1,2$, which act as lowering operators respectively on $\F_\varphi$ and $\F_\Psi$ \footnote{For instance these operators can be those satisfying \eqref{pbn}} in the following way:
\bea
a_1\,\varphi_{n_1,n_2}=\alpha_{n_1}\varphi_{n_1-1,n_2}, \qquad a_2\,\varphi_{n_1,n_2}=\beta_{n_2}\varphi_{n_1,n_2-1},\label{general1}\\
b_1^\dagger\,\Psi_{n_1,n_2}=\alpha_{n_1}\Psi_{n_1-1,n_2}, \qquad b_2^\dagger\,\Psi_{n_1,n_2}=\beta_{n_2}\Psi_{n_1,n_2-1},\label{general2}
\ena
for all $n_j\geq0$. As before, we assume that the norms of the vectors are bounded in a very {\em mild} way:
\bea
\|\varphi_{n_1,n_2}\|&\leq& A_\varphi r_{1,\varphi}^{n_1}r_{2,\varphi}^{n_2}M_{n_1}(1,\varphi)M_{n_2}(2,\varphi),
\label{3add1}\\
\|\Psi_{n_1,n_2}\|&\leq& A_\Psi r_{1,\Psi}^{n_1}r_{2,\Psi}^{n_2}M_{n_1}(1,\Psi)M_{n_2}(2,\Psi),
\ena
for some real constants $A_\Phi$, $r_{k,\Phi}$ and some sequences $M_j(k,\Phi)$, $\Phi$ is both $\varphi$ or $\Psi$, $k=1,2$, $j\geq0$. Then we require that
$$
\lim_{j\rightarrow\infty}\frac{M_j(k,\Phi)}{M_{j+1}(k,\Phi)}=M(k,\Phi),
$$
which can also be divergent. Hence, generalizing Theorem \ref{theo1}, we can define
$$
\rho_1=\overline\alpha\min\left(1,\frac{M(1,\varphi)}{r_{1,\varphi}},\frac{M(1,\Psi)}{r_{1,\Psi}}\right),\quad
\rho_2=\overline\beta\min\left(1,\frac{M(2,\varphi)}{r_{2,\varphi}},\frac{M(2,\Psi)}{r_{2,\Psi}}\right),
$$
and the two related circles $C_{\rho_j}(0)$, $j=1,2$, as well as the following quantities:
\bea
 N(z,w)&=&\left(\sum_{k=0}^\infty\frac{|z|^{2k}}{(\alpha_k!)^2}\right)^{-\frac{1}{2}}\left(\sum_{l=0}^\infty\frac{|w|^{2k}}{(\beta_k!)^2}\right)^{-\frac{1}{2}},
\label{37}\\
\varphi(z,w)&=&N(z,w)\sum_{n_1,n_2\geq0}\frac{z^{n_1}w^{n_2}}{\alpha_{n_1}!\beta_{n_2}!}\,\varphi_{n_1,n_2},\label{38}\\
\Psi(z,w)&=&N(z,w)\sum_{n_1,n_2\geq0}\frac{z^{n_1}w^{n_2}}{\alpha_{n_1}!\beta_{n_2}!}\,\Psi_{n_1,n_2}.
\label{39}\ena
They are all well defined for $z\in C_{\rho_1}(0)$ and $w\in C_{\rho_2}(0)$, and satisfy, for all such $(z,w)$, the normalization condition $\left<\varphi(z,w),\Psi(z,w)\right>=1$. Also:
$$
a_1\varphi(z,w)=z\varphi(z,w), \quad a_2\varphi(z,w)=w\varphi(z,w), $$
and
$$
b_1^\dagger \Psi(z,w)=z\Psi(z,w), \quad b_2^\dagger \Psi(z,w)=w\Psi(z,w).
$$
Concerning the resolution of the identity, this time we have to solve two moment problems: suppose that we can find two measures, $d\lambda_j(r)$, $j=1,2$, such that
$$
\int_0^{\rho_1}d\lambda_1(r)r^{2k}=\frac{(\alpha_k!)^2}{2\pi}, \qquad \int_0^{\rho_2}d\lambda_2(r)r^{2k}=\frac{(\beta_k!)^2}{2\pi}.
$$
for all $k\geq0$. Then, calling $d\nu_1(z,\overline z)=d\lambda_1(r)\,d\theta$ and $d\nu_2(w,\overline w)=d\lambda_1(r')\,d\theta'$, we can prove the following: for all $f,g\in\D$ we have, for instance,
$$
\int_{C_{\rho_1}(0)}d\nu_1(z,\overline z) \int_{C_{\rho_2}(0)}d\nu_2(w,\overline w) N(z,w)^{-2}\left<f,\Psi(z,w)\right>\left<\varphi(z,w),g\right>=
\left<f,g\right>,
$$
and a similar formula with $\Psi(z,w)$ and $\varphi(z,w)$ exchanged.

\vspace{2mm}

{\bf Remark:--} If, in particular, {$\alpha_{n}=\sqrt{n}=\beta_n$, as is the case for the Swanson model, it is clear that $\overline\alpha=\overline\beta=\infty$ and, because of their definitions, $\rho_1=\rho_2=\infty$.  Moreover, $N(z,w)=e^{-\frac{|z|^2+|w|^2}{2}}$ and $\varphi(z,w),\Psi(z,w)$ reduce to \eqref{vphiBCS}-\eqref{vpsiBCS}. This means that convergence of the bi-coherent states  is guaranteed in all $\Bbb C^2$.

 \subsection{Back to Swanson}

 %
%
%
%
%
% To show this we use the following property (see \cite{SIGMA})
%
% \begin{prop}
% Suppose that $r_\nu,t_\nu\geq0$ and $0\leq\alpha_{\nu},\beta_{\nu}\le1/2$ exist such that
% \bea
% \lVert \varphi_{mn}\rVert\leq r_\nu^mt_\nu^n (m!)^{\alpha_{\nu}}(n!)^{\beta_{\nu}},\quad \forall m,n\geq0\label{prop}
% \ena
% then $\varphi(z,w)$ is well defined $\forall z,w \in \mathbb{C}$.
% \end{prop}
% \begin{proof}
% The proof of the above proposition follows easily from the one presented in \cite{SIGMA} and the straightforward factorization for two variables.
% \end{proof}
% Incidentally we observe that, if the operators $\U(z,w)$ and $\V(z,w)$ are well defined, then the vectors in (\ref{BCS}) coincide with those in (\ref{vphiBCS}) and in (\ref{vpsiBCS}).

To apply the previous results to our modified Swanson model we need now to find a relevant estimate for the norms of the vectors in $\F_{\varphi}$ and $\F_{\Psi}$. For that we use the formula (\cite{Pru}, pag. 502):
 \beano
 \int_{0}^{\infty}\int_{0}^{\infty} e^{-p(x^2+y^2)}H_{n_1}(ax)H_{n_1}(bx)H_{n_2}(cy)H_{n_2}(fy)dx dy=\\
\frac{2^{{n_1}+{n_2}-2}{n_1}!{n_2}!\pi}{p^{({n_1}+{n_2}+2)/2}}(a^2+b^2-p)^{{n_1}/2}(c^2+f^2-p)^{{n_2}/2}\times\\
 \times P_{n_1}\left(\frac{ab}{\sqrt{p(a^2+b^2-p)}}\right) P_{n_2}\left(\frac{cf}{\sqrt{p(c^2+f^2-p)}}\right),
 \enano
where $P_n$ is the Legendre polynomial of order $n$. The above formula is valid for all $p$ having non negative real part, and in our context $p=\cos(2\nu)>0, \forall \nu\in I\setminus\{0\}$ \footnote{We still assume that $\nu\neq0$ as we are interested in the non hermitian case}. Straightforward computations finally lead to
\beano
\lVert \varphi_{{n_1,n_2}}\rVert^2=\frac{\pi|N_1|^2}{\cos(2\nu)}P_{n_1}\left(\frac{1}{\cos(2\nu)}\right)P_{n_2}\left(\frac{1}{\cos(2\nu)}\right),
\enano
and using the estimate in \cite{Sze} for $P_n(x)$ we deduce  that
\beano
\lVert\varphi_{n_1,n_2}\rVert\leq A_{\nu}r_\nu^{n_1} t_\nu^{n_2},\quad
r_\nu=t_\nu=\sqrt{\frac{1}{\cos(2\nu)}+\left(\frac{1}{\cos^2(2\nu)}-1\right)^{1/2}},
\enano
with $A_\nu$ a non relevant positive constant. Then, it is clear that the assumption in (\ref{3add1}) is satisfied, taking for instance $M_n(1,\varphi)=M_n(2,\varphi)=1$, for all $n\geq0$.
Similar considerations can be repeated for $\Psi(z,w)$, so that all the results deduced before apply here. In particular $\varphi(z,w)$ are eigenstates of $A_j$, $\Psi(z,w)$ are eigenstates of $B_j^\dagger$ and, solving the above moment problems (which collapse to a single one), they produce a resolution of the identity.

\subsection{What if $\nu$ is complex?}

In the literature on Swanson model, $\nu$ is always taken to be real. We will briefly show now that this is not really essential, at least if its real part still belongs to the set $I$ introduced before. For that, let us assume that $\nu=\nu_r+i\nu_i$, with $\nu\in I$ and $\nu_i\in\Bbb R$. Then, formulas (\ref{21})-(\ref{25}) are still valid. However, (\ref{25bis}) should be replaced with
$$H_{\nu,\theta}^\dagger=\frac{1}{2\cos(\overline{\nu})}\,\left(A_1^\dagger B_1^\dagger +A_2^\dagger B_2^\dagger+\1\right).$$
Also, while the analytical expression of $\varphi_{n_1,n_2}(x_1,x_2)$ in (\ref{2a}) does not change, that of $\Psi_{n_1,n_2}(x_1,x_2)$ can be deduced from $\varphi_{n_1,n_2}(x_1,x_2)$ by replacing $\nu$ with $-\overline{\nu}$. We deduce again that $\varphi_{n_1,n_2}(x_1,x_2)$ and $\Psi_{n_1,n_2}(x_1,x_2)$ are all in $\Sc(\Bbb R^2)$, and therefore in $\Lc^2(\Bbb R^2)$. Also in this extended case, it is possible to check that $\F_\varphi$ and $\F_\Psi$ are not Riesz bases. In fact we find that
\beano
\lVert \varphi_{{n_1,n_2}}\rVert^2=\frac{\pi|N_1|^2}{e^{-2\nu_i}\cos(2\nu_r)}P_{n_1}\left(\frac{1}{\cos(2\nu_r)}\right)
P_{n_2}\left(\frac{1}{\cos(2\nu_r)}\right),
\enano
where $\nu_i$ explicitly appears.
A similar estimate, with $N_1$ replaced by $N_2$, also holds for $\lVert \Psi_{{n_1,n_2}}\rVert^2$. Both these norms diverge when $n_1$ and $n_2$ diverge, see \cite{magnus}. Hence, see \cite{baginbagbook}, $\F_\varphi$ and $\F_\Psi$ cannot be Riesz bases, also for complex $\nu$. For this reason, no major differences are expected with respect to our previous results.

\section{Conclusions}

In this paper we have proposed a non-commutative, two-dimensional, version of the Swanson model and we have shown that its Hamiltonian can be rewritten in terms of $\D$-pseudo-bosonic operators. In this way, the eigenvalues and the eigenvectors can be easily deduced. We have also considered the bi-coherent states attached to the model, analyzing some of their properties. In particular, the fact that they resolve the identity has been proved.

\section*{Acknowledgements}

F.B. and F.G. acknowledges support by the University of Palermo and GNFM of INdAM.

\end{document}